\documentclass{jps-cp}
\usepackage{txfonts} 

\newcommand{\be}{\begin{equation}}
\newcommand{\ee}{\end{equation}}
\newcommand{\beq}{\begin{eqnarray}}
\newcommand{\eeq}{\end{eqnarray}}

\renewcommand{\>}{\rangle}

\title{Hadron spin structure from lattice QCD}

\author{Constantia \textsc{Alexandrou}}

\inst{Department of Physics, University of Cyprus, PO Box 20537, 1678 Nicosia, Cyprus,\\ and\\
  Computation-based Science and Technology Research Center, The Cyprus Institute, 20, Constantinou Kavafi Str., Nicosia 2121, Cyprus}

\email{alexand@ucy.ac.cy}

\recdate{November  30, 2021} 

\abst{We present results on the spin carried by quarks and gluons in  the nucleon  using lattice QCD simulations with physical values of the light, strange and charm quark masses. We also discuss selective results on the  x-dependence of parton distribution functions computed in lattice QCD employing the quasi-parton distribution approach and the large momentum effective theory. }

\kword{Lattice QCD, hadron structure, spin structure of hadrons}

\begin{document}
\maketitle

\section{Introduction}

Quantum Chromodynamics (QCD) is the fundamental theory of the strong interactions that describes a plethora of complex nuclear physics phenomena including giving rise to hadron states, formed by quarks and gluons.   The most fundamental hadron is the proton, which, being a stable particle, has served for many years   as a laboratory for studying  the complex dynamics of QCD, both theoretically and experimentally. A key  question that has been posed for many years is how the spin of the proton  arises  from its constituents, the quarks and the gluons, collectively called partons. The successful quark model, that describes well properties of the low-lying hadrons, predicted that all the spin is carried by the three valence quarks.
A major surprise came from the measurements of the European Muon Collaboration (EMC)~\cite{EuropeanMuon:1987isl,EuropeanMuon:1989yki} that determine  the proton spin-dependent structure function down to momentum fraction $x=0.01$. The conclusion was that only about half  of the proton spin is carried by the valence quarks. This came to be known as the proton spin puzzle.  It triggered a series of precise measurements by the Spin Muon Collaboration (SMC) in 1992-1996~\cite{SpinMuonSMC:1993coa} and by COMPASS~\cite{Ball:2003vb} since 2002.  Recent experiments using polarized deep inelastic lepton-nucleon scattering (DIS) processes indeed confirmed that only about 25-30\%~\cite{deFlorian:2008mr,deFlorian:2009vb,Blumlein:2010rn,Leader:2010rb,Ball:2013lla,Deur:2018roz} of the nucleon spin comes from the valence quarks.

\begin{figure}[h!]
  \begin{minipage}{0.48\linewidth}
     \includegraphics[width=\linewidth]{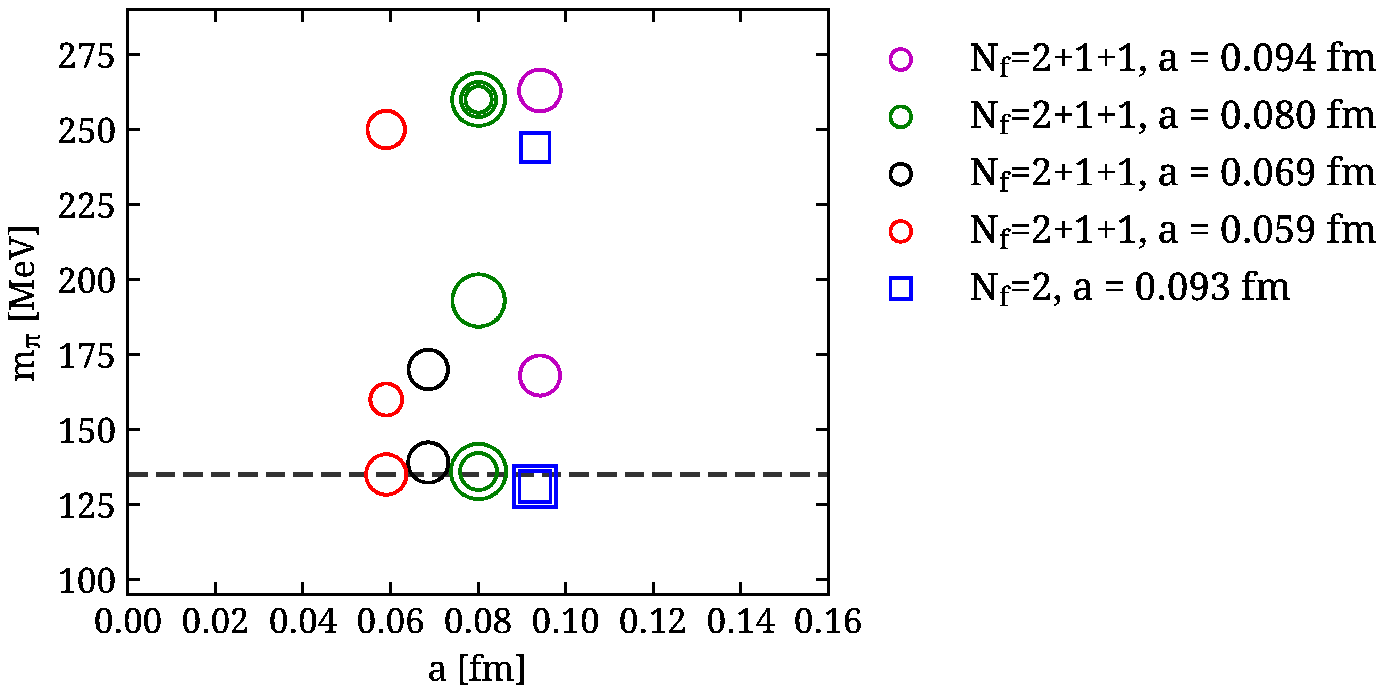}
  \end{minipage}\hfill
  \begin{minipage}{0.48\linewidth}
    \center{ \includegraphics[width=\linewidth]{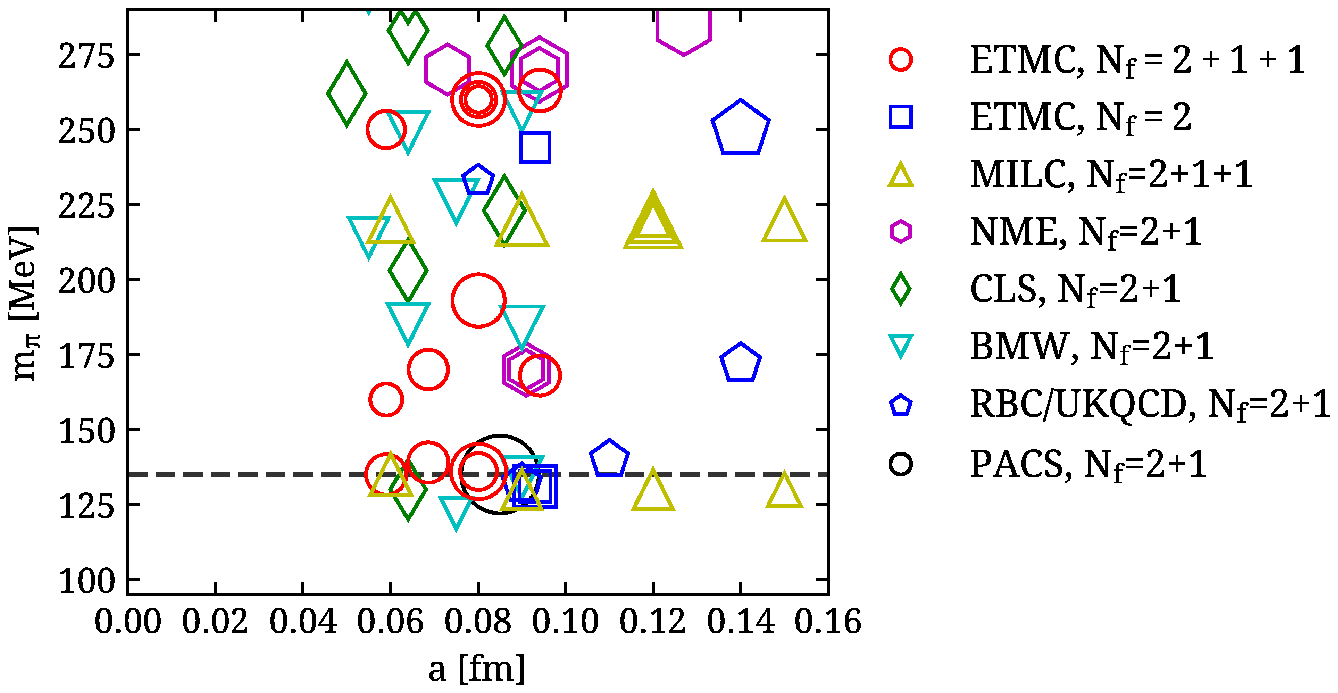}}
  \end{minipage}
\caption{Left: Ensembles of TMF configurations generated with a mass degenerate doublet of up and down quarks, referred to as light quarks and denoted as  $N_f=2$ ensembles, as well as with  a strange and charm quark ($N_f=2+1+1$),  as the mass of the light quarks is varied and as function of the lattice spacing $a$. The mass of the strange and charm quarks are tuned to approximately their physical values. The  size of the symbol is proportional to  the spatial extent of the lattice. Right: A summary of zero-temperature simulations that include clover, twisted mass, staggered and domain
    wall fermions by various lattice QCD collaborations, as indicated in the legend.}
\label{fig:simulations}\vspace*{-0.7cm}
\end{figure}

While experiments play a crucial role in the understanding of the origin of the proton spin, a theoretical approach to compute the various contributions is equally important. 
However, due to the non-perturbative nature of QCD, computing the parton contributions to the spin is difficult. See  Ref.~\cite{Ji:2020ena} for a recent theoretical overview on the nucleon spin. 
Lattice QCD provides the {\it initio} non-perturbative framework that is  suitable to address the key question of how the nucleon spin and  momentum are distributed among its constituents using directly the QCD Lagrangian. Tremendous progress has been made in simulating lattice QCD in recent years. State-of-the-art simulations are currently being performed with   dynamical up, down and strange quarks with masses tuned to their physical values.  A subset of simulations also include a dynamical  charm quark with mass fixed to approximately its physical value. A particularly suitable discretization scheme for studying hadron structure is the so called twisted mass fermion (TMF) action employed by the Extended Twisted Mass Collaboration (ETMC).  This progress was made possible using  efficient algorithms and in particular multigrid solvers~\cite{Clark:2016rdz} that were developed for twisted mass fermions~\cite{Alexandrou:2018wiv}. In Fig.~\ref{fig:simulations} we show the ensembles generated by ETMC as well as the landscape of currently available gauge ensembles worldwide. As can be seen, ETMC and a number of other collaborations have now ensembles generated with physical values of the pion mass, referred to as physical point ensembles. We also note that currently all ensembles are generated with lattice spacing $a>0.05$~fm due to the so-called critical slow down of simulations. New approaches that include machine learning  are being explored for addressing this issue.  

\begin{figure}[t!]
 \hspace*{-0.7cm}{\includegraphics[width=\linewidth]{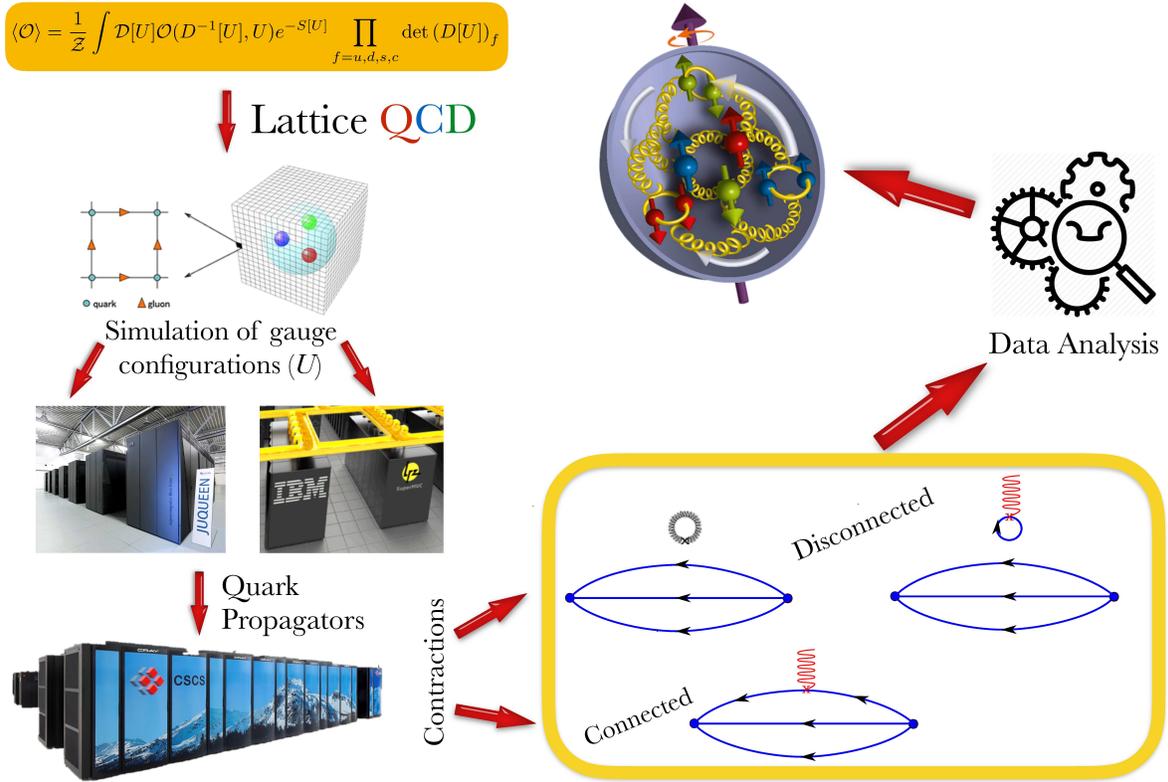}}
    \caption{A typical work-flow for a hadron structure computation within lattice QCD.
   $S(U)$ is the action in Euclidean time and the integration is over the gauge links $U$. The expectation value of an appropriately defined operator ${\cal{O}}$ yields the physical properties we are interested in.  The diagrams shown in the lower left box correspond to the connected and disconnected three-point correlators needed for baryon
    structure studies.}
    \label{fig: workflow}\vspace*{-0.7cm}
\end{figure}

Although lattice QCD provides an exact formalism for solving QCD, one needs a careful study of systematic errors.
In the past, the absence of simulations at physical values of the light quark mass necessitated a chiral
extrapolation.
For the pion sector such an extrapolation using e.g. NLO SU(2) chiral perturbation theory works well for pion masses
$m_\pi\stackrel{<}{\sim}250$~MeV~\cite{Durr:2014oba}.
In the nucleon sector, chiral extrapolation is more problematic and introduces an uncontrolled systematic
error~\cite{Colangelo:2010ba}.
Nowadays, with simulations with physical pion mass this systematic error can be eliminated.
Therefore, in what follows we will focus on results obtained using simulation generated with physical pion mass, since we will focus on nucleon
properties.
Other systematic effects that need to be investigated are:
\begin{itemize}
\item Discretization effects: Since the computation is done at finite lattice spacing $a$ one needs to take the continuum limit.
This requires simulations with at least three values of $a$, ideally at fixed quark masses and volume.
\item Volume effects: A numerical evaluation is necessarily done using a finite volume.
At least three volumes would be required ideally at fixed quark masses and $a$ to take the infinite volume limit.
\item Renormalization: Matrix elements computed on the lattice must be properly renormalized in order to compare with what is
measured in the laboratory.
In state-of-the-art lattice QCD computations, renormalization is carried out nonperturbatively.
However, for gluon and flavor singlet quantities there is mixing and nonperturbative renormalization is still difficult and
a subject of on-going study.
\item Extraction of nucleon matrix elements: Extracting the ground state from a tower of higher excited states necessitates taking
the large Euclidean time limit that gives rise to large gauge noise that makes difficult the identification of ground state properties and requires very large statistics. The development of noise reduction techniques is thus an important ongoing effort.
\end{itemize}
The work-flow of a typical lattice QCD computation for baryon structure is shown schematically in Fig.~\ref{fig: workflow}. After defining the theory on a discrete finite 4-dimensional Euclidean lattice, one generates representative configurations of gauge links $U(x)_\mu$ via a Markov Chain Monte Carlo. 
The three-point correlation functions can be classified as  connected when a probe couples to a valence quark and as disconnected when it
couples to a sea quark or a gluon as shown in Fig.~\ref{fig: workflow}.
Disconnected quark loop contributions require special techniques and larger computational resources. The generation of $U_\mu(x)$  together with the inversion of the Dirac matrix $D[U]$ to produce the quark propagator, are the most time comsuming parts of the calculation. While the action $S[U]$ is local, $\rm {det(D[U])}$ is not and it took years of algorithmic developments to be able to efficiently include it in the simulations. There are several discretization schemes for $D{U}$, each having its pros and cons. The most widely used are clover (employed e.g. by BMW, CLS, and PACS) and twisted mass fermions (ETMC), overlap and domain wall fermions (RBC/UKQCD) and   staggered fermions (MILC)  giving rise to the different collaborations mentioned in Fig.~\ref{fig:simulations}. All schemes coincide in the continuum limit.

\section{Spin structure in lattice QCD}

In lattice QCD studies, one typically starts from the decomposition of the energy-momentum tensor (EMT) into traceless and trace
parts~\cite{Ji:1995sv}:
\be
    T^{\alpha\nu} = \bar{T}^{\alpha\nu} + \hat{T}^{\alpha\nu}.
    \label{EMT-decompostion}
\ee
The traceless part, $\bar{T}^{\alpha\nu}$, decomposes into two scheme and scale dependent terms,
\be
\bar{T}^{\alpha\nu} = \bar{T}_q^{\alpha\nu}(\mu) + \bar{T}_g^{\alpha\nu}(\mu),
\label{traceless}
\ee
given by
\be
\bar{T}_q^{\alpha\nu}(\mu) = \bar{\psi} i \gamma^{\{\alpha} \overleftrightarrow{D}^{\nu\}} \psi \hspace*{0.4cm}
{\rm and} \hspace*{0.4cm}\bar{T}_g^{\alpha\nu}(\mu) = F^{\{ \alpha \rho} F^{\nu\}}_{\;\;\;\;\rho},
\label{EMT-decomposition2}
\ee where $\mu$ is the energy scale and curly braces mean symmetrization over indices \emph{and} subtraction of
trace. These are related, respectively, to  the quark and gluon total angular momentum.

\subsection{Nucleon matrix elements}

In order to compute the nucleon spin, we need to evaluate  the nucleon matrix elements of the EMT. They  can be decomposed into generalized form factors (GFFs) that depend only on the momentum transfer squared $q^2$. In Minkowski space we have~\cite{Ji:1998pc}
\be
   \langle N(p',s') \vert \bar{T}^{\mu \nu}_{q,g} \vert N(p,s) \rangle = \bar{u}_N(p',s') \bigg[
    A_{20}^{q,g}(q^2) \gamma^{\{\mu} P^{\nu\}}
    + B_{20}^{q,g}(q^2) \frac{ i \sigma^{\{\mu \rho} q_\rho P^{\nu\}} }{2 m_N} 
    + C_{20}^{q,g}(q^2) \frac{q^{\{ \mu} q^{\nu\}}}{m_N}\bigg] u_N(p,s)
    \label{Eq:Decomp}
\ee
where $u_N$ is the nucleon spinor with initial (final) momentum $p(p')$ and spin $s(s')$,  $P=(p'+p)/2$ is the total momentum and $q=p'-p$ the momentum transfer.  $A_{20}^{q,g}(q^2)$, $B_{20}^{q,g}(q^2)$ and $C_{20}^{q,g}(q^2)$ are the three GFFs.
In the forward limit, $A_{20}^{q,g}(0)$ gives the quark and gluon average momentum fraction $\langle x \rangle^{q,g}$. Summing over all quark and gluon contributions gives the momentum sum  $\langle x \rangle^{q} + \langle x \rangle^g = 1$. Furthermore,  the total spin carried by a quark is given by $J^q=\frac{1}{2}\left[A^q_{20}(0)+B^q_{20}(0)\right]$~\cite{Ji:1996ek}.

The lattice operator $\bar{T}_{q,g}^{\mu\nu}$ must be renormalized
before we can connect it to physical matrix elements.
For the flavor nonsinglet combination, this operator is multiplicatively renormalized.
The standard approach to compute the nonsinglet renormalization function $Z_q$ nonperturbatively is to use an analysis within
a regularization indepedent (RI) scheme. 
The flavor singlet operator, however, mixes with the gluon operator and  one needs to compute a $2\times2$ matrix element.
The mixing coefficients ($Z_{qg}$, $Z_{gq}$) are thus needed to extract the momentum fraction and spin carried by quarks and gluons.
The $2\times2$ mixing matrix needed for the proper renormalization procedure of the momentum fractions is given by
\be
\langle x \rangle_R^{q^+} =  Z_{qq} \,\langle x \rangle_B^{q^+} + Z_{qg}\, \langle x \rangle_B^g, \hspace*{1cm}
\langle x \rangle_R^{g}   =  Z_{gg} \,\langle x \rangle^B_{g}   + Z_{gq}\, \langle x \rangle_B^{q^+} .
\label{eq:xgR}
\ee
In the above equations, $\langle x \rangle^{q^+}$ is understood to be the flavor singlet combination that sums the up, down, strange
and charm quark contributions.
The subscript $B$ represents the bare matrix elements.
We note that a complete calculation of the $2 \times 2$ mixing matrix would require the solution of a system of four coupled
renormalization conditions that involve vertex functions of both gluon and quark EMT operators.
One can impose decoupled renormalization conditions for the nonperturbative calculation of the diagonal elements, namely $Z_{gg}$
and $Z_{qq}$, since the mixing coefficients $Z_{gq}$ and $Z_{qg}$ are small in one-loop perturbation theory~\cite{Alexandrou:2016ekb,Alexandrou:2020sml}. The computation of $Z_{qq}$ proceeds analogously to the  non-singlet case.
The computation of the renormalization function $Z_{gg}$ of the gluon operator is more challenging due to the high noise-to-signal
ratio appearing in the calculation of gluonic quantities.
To this end, some equivalent renormalization prescriptions have been proposed to reduce the statistical
uncertainties in the extraction of $Z_{gg}$.
Details on the different renormalization approaches of the gluonic operator are given in
Refs.~\cite{Shanahan:2018pib,Yang:2018nqn,Alexandrou:2020sml}.
The off-diagonal elements  so far are only computed in perturbation theory.

\subsection{Extraction of nucleon matrix elements}
In order to compute the nucleon matrix elements of the previous section, one
needs to evaluate two- and three-point functions in Euclidean space. The 
nucleon two-point function is given by
\be
C(\Gamma_0,\vec{p};t_s,t_0) {=}  \sum_{\vec{x}_s} \hspace{-0.1cm} e^{{-}i (\vec{x}_s{-}\vec{x}_0) \cdot \vec{p}} \times \tr \left[ \Gamma_0 {\langle}{\cal J}_N(t_s,\vec{x}_s) \bar{\cal J}_N(t_0,\vec{x}_0) {\rangle} \right],
\label{Eq:2pf}
\ee
where ${\cal J}_N(t,\vec{x})$ is the nucleon interpolator, $x_0$ is the initial lattice site at which states with the quantum numbers of the nucleon are created, referred to as source position and $x_s$ is the site where they are annihilated, referred to as sink. The three-point function is given by
\be
  C^{\mu\nu}(\Gamma_\rho,\vec{q},\vec{p}\,';t_s,t_{\rm ins},t_0) {=}
 \hspace{-0.1cm} {\sum_{\vec{x}_{\rm ins},\vec{x}_s}} \hspace{-0.1cm} e^{i (\vec{x}_{\rm ins} {-} \vec{x}_0)  \cdot\vec{q}}  e^{-i(\vec{x}_s {-} \vec{x}_0)\cdot \vec{p}\,'} {\times}  \tr \left[ \Gamma_\rho \langle {\cal J}_N(t_s,\vec{x}_s) \bar{T}_{q,g}^{\mu\nu}(t_{\rm ins},\vec{x}_{\rm ins}) \bar{\cal J}_N(t_0,\vec{x}_0) \rangle \right].
  \label{Eq:3pf}
\ee
 The Euclidean momentum transfer squared is given $Q^2=-(p'-p)^2$ and  $\Gamma_\rho$ is the projector acting on the spin indices. We consider $\Gamma_0=\frac{1}{2}(1+\gamma_0)$ and $\Gamma_{k}=i\Gamma_0 \gamma_5 \gamma_k$ taking the non-relativistic representation for $\gamma_\mu$.

 In order to extract the desired nucleon matrix element, we construct appropriate combinations of three- to two-point functions, which in the large Euclidean time limit, cancel the time dependence arising from the  time propagation and the overlap terms between the interpolating field and the nucleon state. An optimal choice that benefits from  correlations is the ratio~\cite{Alexandrou:2013joa,Hagler:2003jd}.
\be
 R^{\mu\nu}(\Gamma_{\rho},\vec{p}\,',\vec{p};t_s,t_{\rm ins}) = \frac{C^{\mu\nu}(\Gamma_\rho,\vec{p}\,',\vec{p};t_s,t_{\rm ins}\
)}{C(\Gamma_0,\vec{p}\,';t_s)} \times   \sqrt{\frac{C(\Gamma_0,\vec{p};t_s-t_{\rm ins}) C(\Gamma_0,\vec{p}\,';t_{\rm ins}) C(\Gamma_0,\vec{p}\,';t_s)}{C\
(\Gamma_0,\vec{p}\,';t_s-t_{\rm ins}) C(\Gamma_0,\vec{p};t_{\rm ins}) C(\Gamma_0,\vec{p};t_s)}}.
\label{Eq:ratio}
\ee
The sink and insertion time separations  $t_s$ and $t_{\rm ins}$ are taken relative to the source time $t_0$. Taking the limits $(t_s-t_{\rm ins}) \gg a$ and $t_{\rm ins} \gg a$ in 
Eq.(\ref{Eq:ratio}), filters out lowest nucleon state. When this happens, the ratio
becomes independent of time
\begin{equation}
  R^{\mu\nu}(\Gamma_\rho;\vec{p}\,',\vec{p};t_s;t_{\rm ins})\xrightarrow[t_{\rm ins}\gg a]{t_s-t_{\rm ins}\gg a}\Pi^{\mu\nu}(\Gamma_\rho;\vec{p}\,',\vec{p})\,
\end{equation} 
and yields the desired nucleon matrix element.
In practice,   $(t_s-t_{\rm ins})$ and $t_{\rm ins} $ cannot be taken arbitrarily large, since the signal-to-noise ratio decays exponentially  with the sink-source time separation. Therefore, one needs to take  $(t_s-t_{\rm ins})$ and $t_{\rm ins}$ large enough so that the nucleon state dominates in the ratio. To identify when this happens is  a delicate process and we use the following  three methods:

\noindent \emph{Plateau method:} 
If contributions from  excited states are sufficiently suppressed, the ratio of Eq.~(\ref{Eq:ratio}) can be written as 
  \be
  \Pi^{\mu\nu}(\Gamma_\rho;\vec{p}\,',\vec{p}) + {\cal O}(e^{-\Delta E(t_s - t_{\rm ins})}) + {\cal O}(e^{-\Delta E t_{\rm ins}})+\cdots,
  \ee
where  $\Delta E$ is the energy gap between the nucleon state and the
first excited state. The first  time-independent term is the matrix element we interested in. Thus, 
  we seek to identify  nucleon state dominance by  looking  for a range of values of  $t_{\rm ins}$ for which  the ratio of Eq.~(\ref{Eq:ratio}) is time-independent (plateau region). 
We fit the ratio to a constant within the  plateau region and seek to see convergence in the extracted fit values as
we increase $t_s$. If such a  convergence can be demonstrated, then  the desired nucleon matrix element can be extracted.

\noindent \emph{Summation method:}  One can sum over $t_{\rm ins}$ the ratio of Eq.~(\ref{Eq:ratio}) to obtain
\be
  R^{\mu\nu}_{\rm summed}(\Gamma_\rho;\vec{p}\,',\vec{p};t_s) = \sum_{t_{\rm ins}=2a}^{t_s-2a} R^{\mu\nu}(\Gamma_\rho;\vec{p}\,',\vec{p}\
;t_s;t_{\rm ins}) = 
  c + \Pi^{\mu\nu}(\Gamma_\rho;\vec{p}\,',\vec{p}) {\times}t_s + {\cal O}(e^{- \Delta E t_s}).
\label{Eq:Summ} 
\ee
Assuming the nucleon state dominates over excited state contributions,  the desired matrix element given by $\Pi_{\mu\nu}(\Gamma_\rho;\vec{p}\,',\vec{p})$ is extracted from the slope of a linear fit with respect to $t_s$. As in the case of the plateau method,  we probe convergence by increasing the lower  value of $t_s$used in the fit until the resulting values for the slope converge. While both plateau  and summation methods assume that the  ground state dominates, the exponential suppression of excited states in the summation is faster and  approximately corresponds to using twice the sink-source time separation $t_s$ in the plateau method.

\noindent \emph{Two-state fit method:}
One can  explicitly include the contributions from the first excited state and fit. The expressions and details on the analysis  are given in Ref.~\cite{Alexandrou:2020sml}.

\section{Results on the spin decomposition of the nucleon}

\begin{figure}[ht!]
  \begin{minipage}{0.48\linewidth}
  \includegraphics[width=\linewidth]{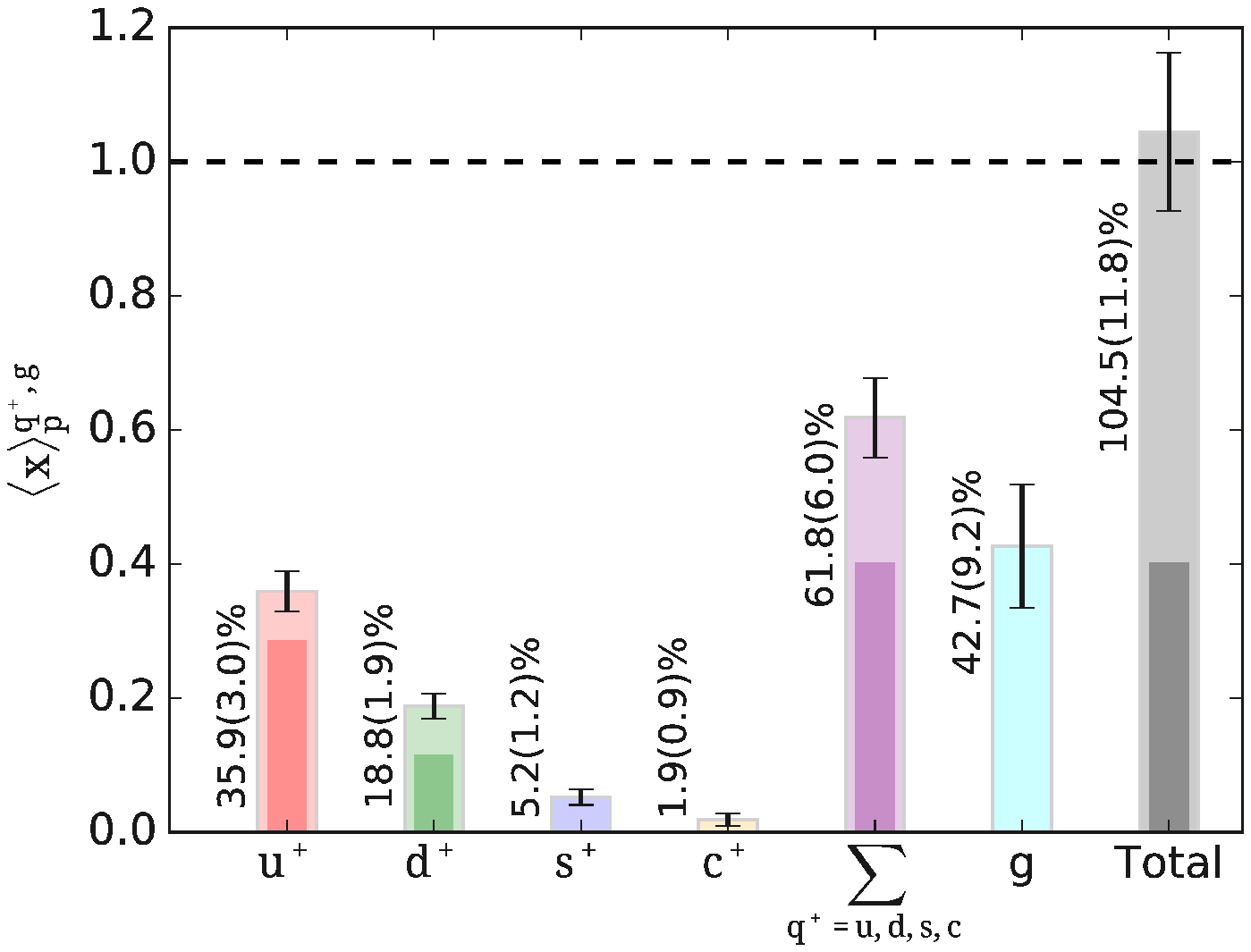}
  \end{minipage}\hfill
    \begin{minipage}{0.48\linewidth}
  \includegraphics[width=\linewidth]{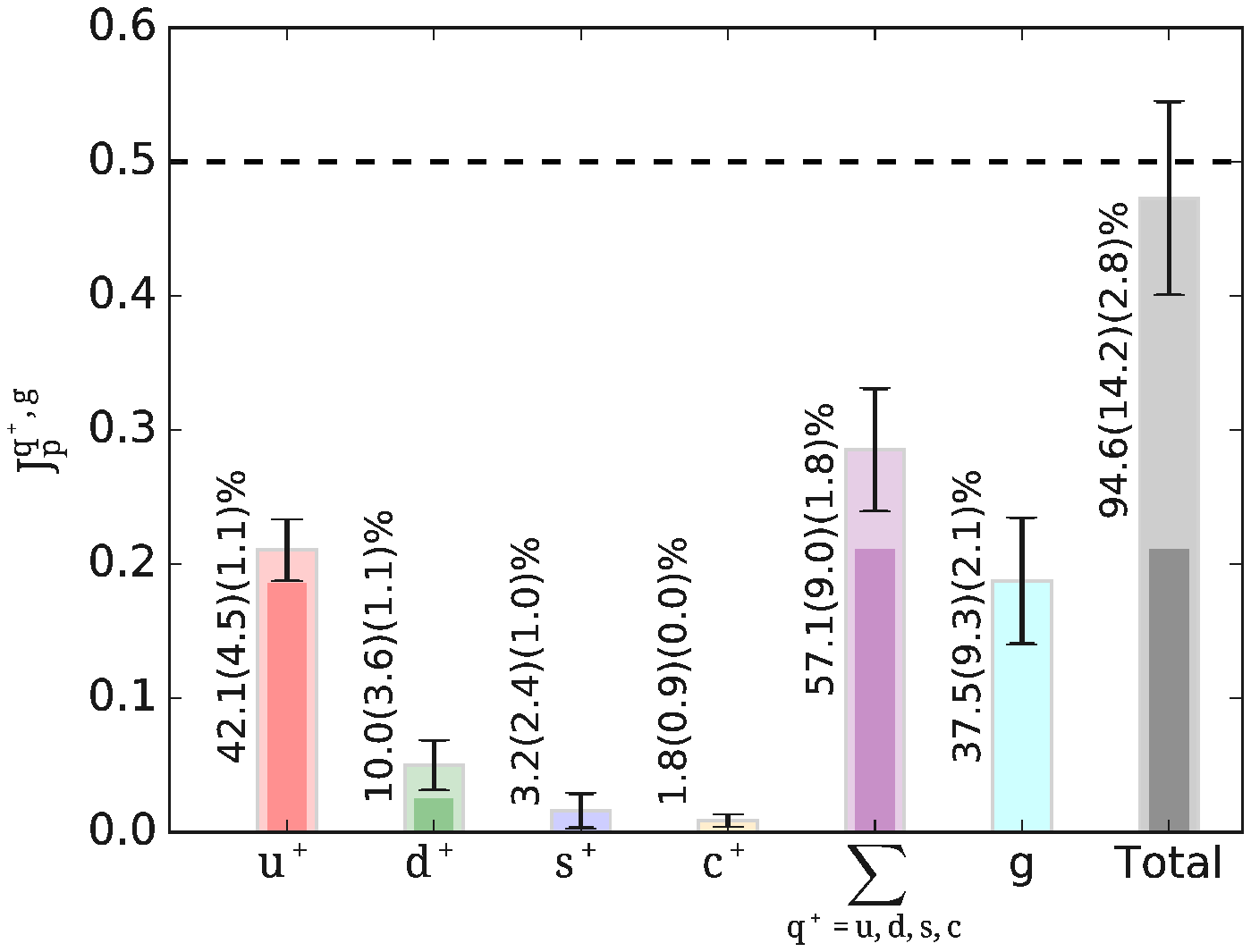}
\end{minipage}
 \caption{The decomposition of the proton average momentum fraction $\langle x \rangle_P$ (left) and  spin $J_P$. We show the contribution of  the up (red bar), down (green bar), strange (blue bar) and charm (orange bar) quarks and their sum (purple bar),  the gluon (cyan bar) and the total sum (grey bar). Note that what is shown  is the contribution of both the  quarks and antiquarks ($q^+=q+\bar{q}$). 
 Whenever two overlapping bars appear the darker bar denotes the purely connected contribution while the light one is the total contribution, which includes disconnected taking into account also the mixing. The error bars on the only connected part are omitted while 
 for the total are shown explicitly on the bars. The percentages written in the figure are for the total contribution.
 The dashed horizontal line is the momentum and spin sums. Results are given in $\mathrm{ \overline{MS}}$ scheme at 2~GeV.}\vspace*{-0.7cm}
  \label{fig:averXBar}
 \end{figure}

After determining the  nucleon  matrix element from the ratio of  Eq.~(\ref{Eq:ratio} we
can extract nucleon charges, moments and  GFFs.
In Fig.~\ref{fig:averXBar}, we show  our results for the proton average momentum
fraction for the up, down, strange and charm quarks, for the gluons as well as their sum.
The up quark makes the largest quark contribution of about 35\% and it is twice as big as that of the down
quark. The strange quark contributes significantly smaller, namely about 5\% and the charm contributes about 2\%.
The gluon has a significant
contribution of about 45\%.  Summing all the contributions 
results to
\be
\sum_{q=u,d,s,c}\langle x \rangle_R^{q^+} + \<x\>_R^g = 104.5(11.8)
\ee
confirming the expected momentum sum.
Fig.~\ref{fig:averXBar}, showing connected and disconnected contributions,  demonstrates that disconnected contributions are crucial and if  excluded 
would result to a significant underestimation of the momentum sum.  

The individual contributions to the proton spin are presented in Fig.~\ref{fig:averXBar}. The major contribution comes from the up quark amounting to about 40\% of the proton spin. The down, strange and charm quarks have relatively smaller contributions. All quark flavors together constitute to about 60\% of the proton spin. The gluon contribution is as significant as that of the up quark,  providing the missing piece to obtain $J_P=94.6(14.2)(2.8)$\%  of the proton spin, confirming indeed the spin sum.
The $\sum_{q=u,d,s} B_{20}^{q^+}(0) + B_{20}^g(0)$ is expected to vanish in order to respect the momentum and spin sums. We find for the renormalized values that $
    \sum_{q=u,d,s} B_{20,R}^{q^+}(0) + B_{20,R}^g(0)=-0.099(91)(28)$,
which is indeed compatible with zero.

\begin{figure}[ht!]
   \begin{minipage}{0.48\linewidth}
  \includegraphics[width=\linewidth]{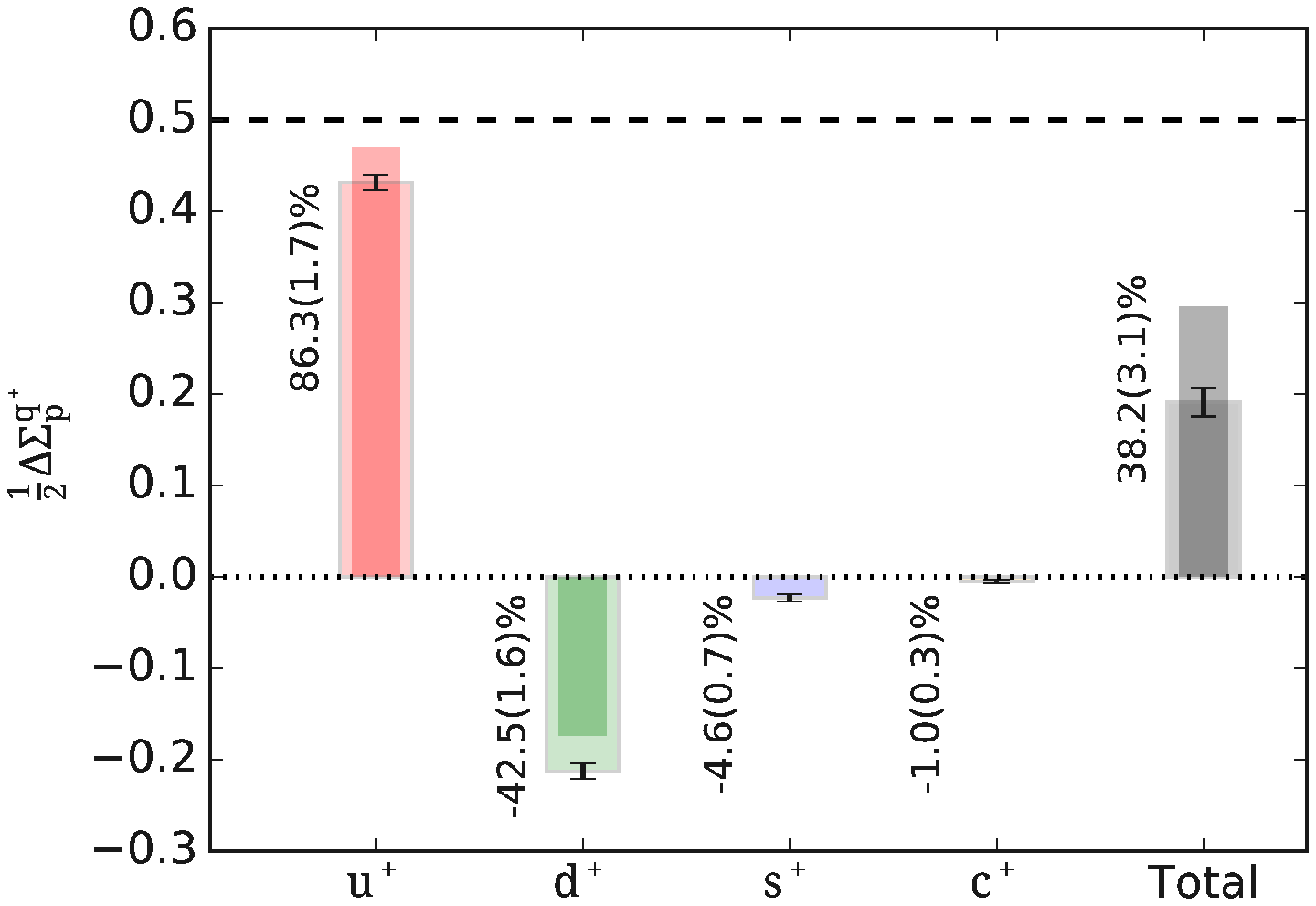}
  \end{minipage}\hfill
    \begin{minipage}{0.48\linewidth}
  \includegraphics[width=\linewidth]{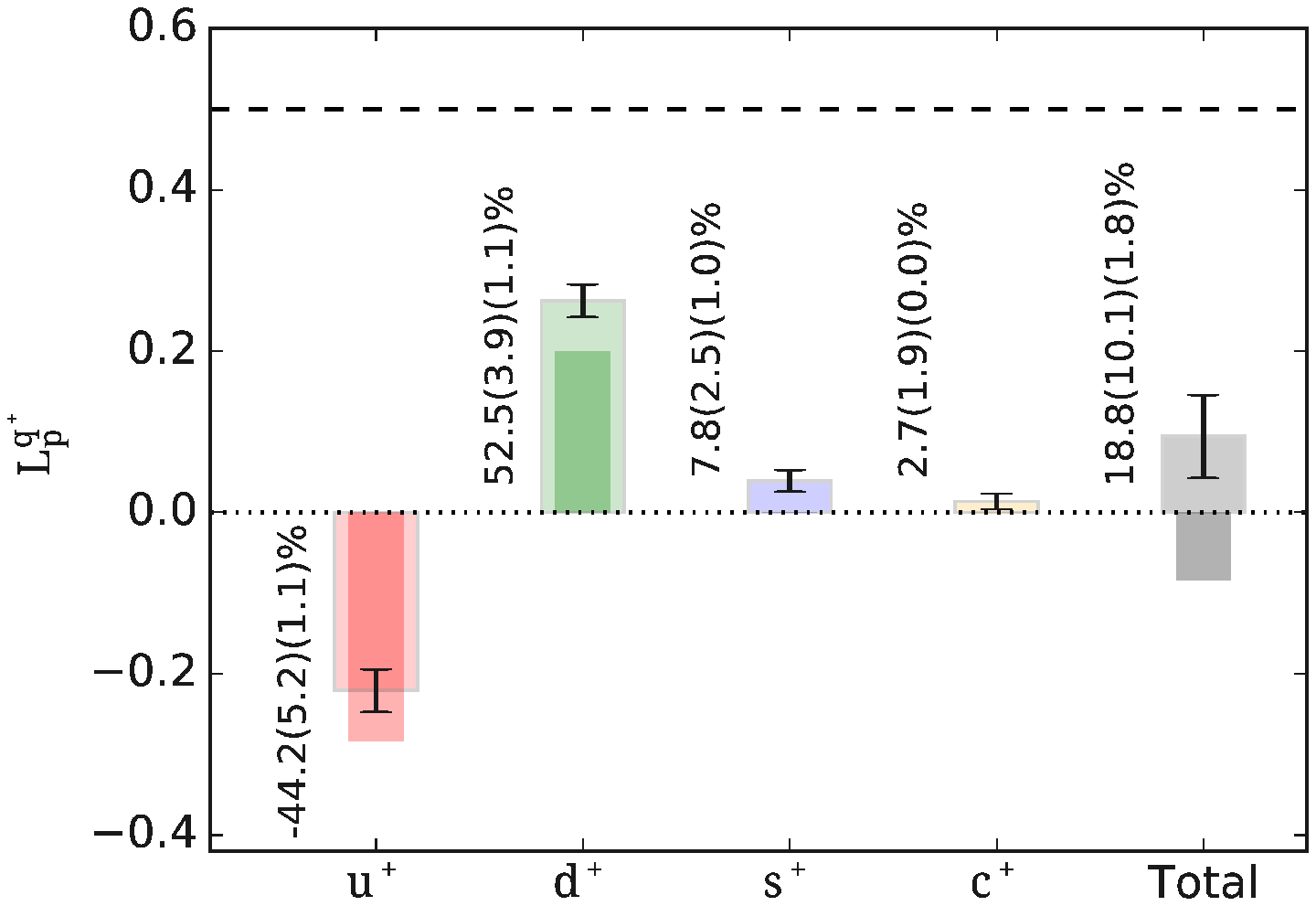}
\end{minipage}
 \caption{Results  for the intrinsic quark spin $\frac{1}{2} \Delta \Sigma$ (left) and orbital angular momentum $L$ (right). We given the contributions to the proton spin decomposed into the  up (red bar), down (green bar), strange (blue bar), and charm (orange bar) quark parts. The total contribution of the four flavors is also shown (grey bar)~\cite{Alexandrou:2019brg}.  The dashed horizontal line is the observed proton spin and the percentages are given relative to it. Results in $\mathrm{ \overline{MS}}$ scheme at 2~GeV.}\vspace*{-0.7cm}
 \label{fig:DeltaSigmaBar}
 \end{figure}

Since the quark contribution to the proton spin is computed, it is interesting to see how much comes from the 
intrinsic quark spin. In Fig.~\ref{fig:DeltaSigmaBar} we show our results for the intrinsic quark spin  $\frac{1}{2} \Delta \Sigma^{q^+}=\frac{1}{2}g_A^{q^+}$, where $g_A^q$ is the axial charge for each quark, the values of which  
 are taken from Ref.~\cite{Alexandrou:2019brg}. The up quark has a large contribution of about 85\% of the proton intrinsic  spin. The down quark contributes about half compared to the up and with  opposite sign. The strange and charm quarks also contribute  negatively  with the latter being about five times smaller than the former giving   a 1\% contribution.
The total $\frac{1}{2} \Delta \Sigma^{q^+}$ is in agreement with the upper bound from COMPASS~\cite{COMPASS:2015mhb}
 
Having both the quark total angular momentum and the quark intrinsic spin allows us to extract the orbital
angular momentum $L_P^{q^+}$. 
Our results are shown in Fig.~\ref{fig:DeltaSigmaBar}. The orbital angular 
momentum of the up quark is negative reducing the total angular momentum contribution  of the up quark to the proton spin. The contribution of  the down quark to
the orbital angular momentum is positive almost canceling the negative intrinsic spin contribution resulting to a 
relatively small positive contribution to the spin of the proton. For a direct calculation of $L_P$ using TMDs see Ref.~\cite{Engelhardt:2017miy}.
  
\section{Direct computation of x-dependencne of parton distributions}
While for many years only GFFs were accessible in lattice QCD, a new approach was proposed that enables us to compute the $x$-dependence of parton distribution functions (PDFs) within lattice QCD taking advantage of the so-called large momentum effective theory (LaMET)~\cite{Ji:2013dva}. The main idea is to compute spatial correlators in lattice QCD  e.g. along $z$-axis and boost the nucleon state in the same direction to large momentum. After renormalization, a matching kernel computed perturbatively is used to recover the light-cone correlation matrix element. Here we will present results using the quasi-PDF approach for which there have been many studies. For recent reviews see Refs.~\cite{Lin:2017snn,Constantinou:2020hdm,Ji:2020ect,Cichy:2021ewm}.
\begin{figure}[ht!]
   \begin{minipage}{0.33\linewidth}
     \includegraphics[width=\linewidth]{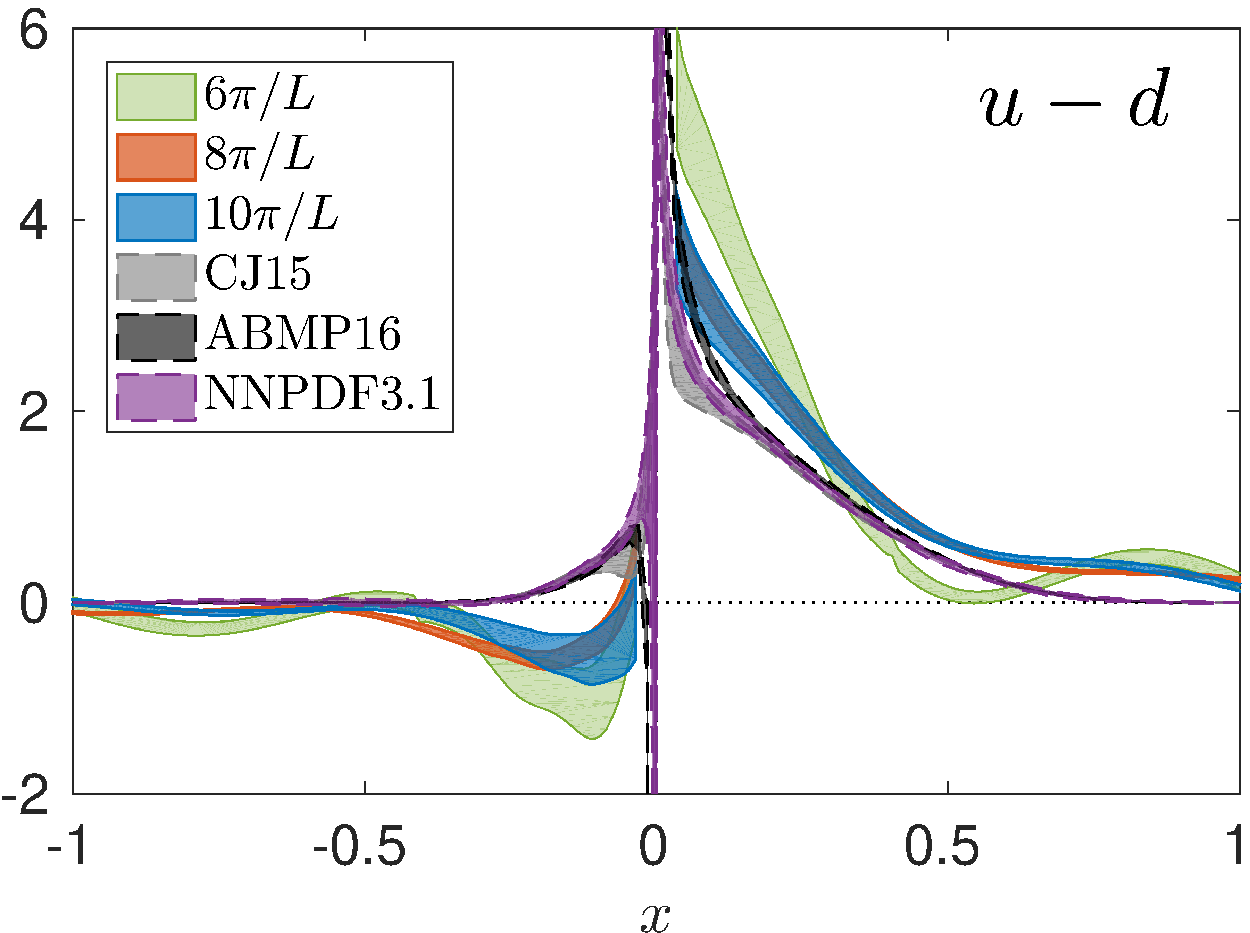}
   \end{minipage}\hfill
      \begin{minipage}{0.33\linewidth}
     \includegraphics[width=\linewidth]{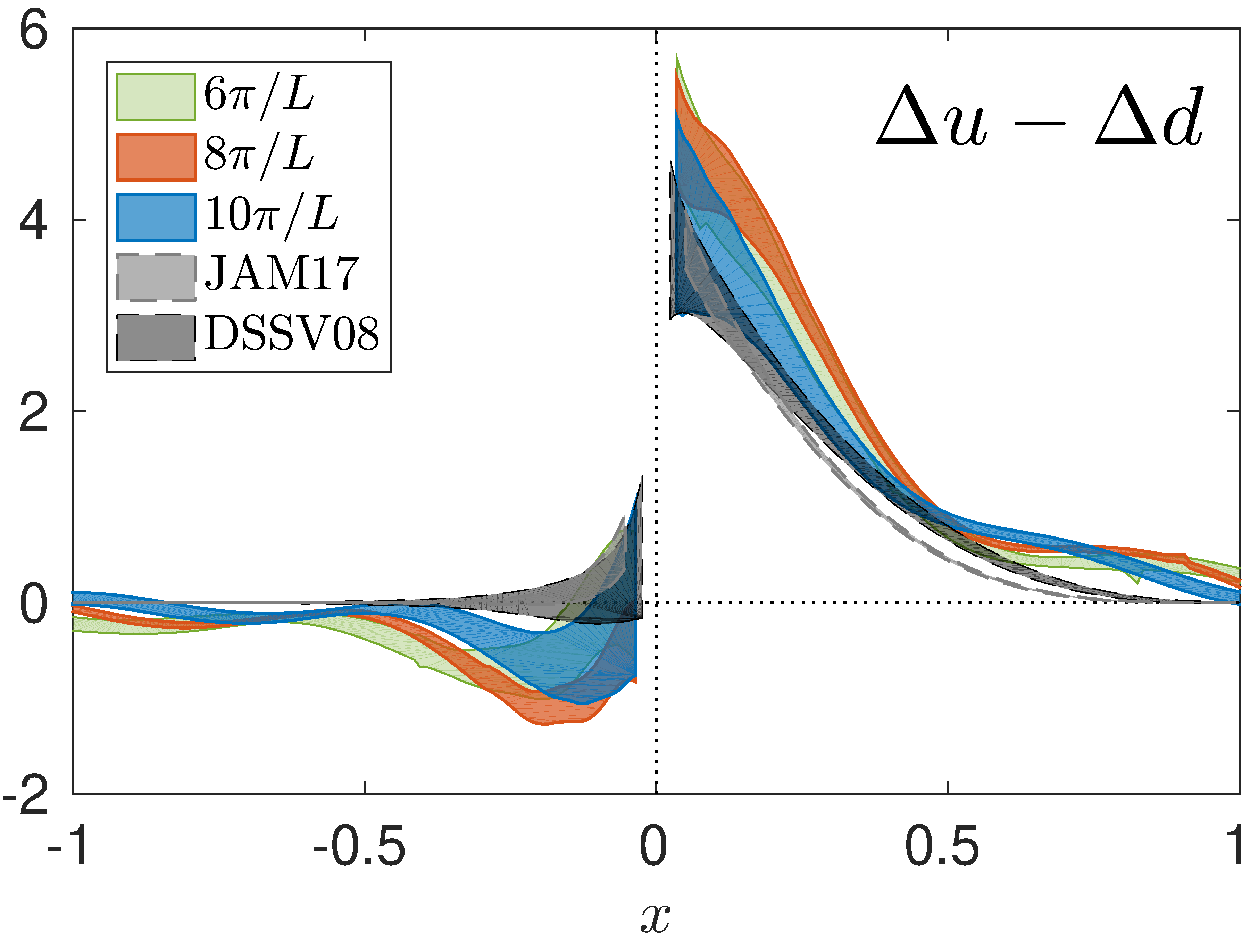}
   \end{minipage}\hfill
   \hspace{-0.3cm}\begin{minipage}{0.31\linewidth}
 \includegraphics[width=\linewidth]{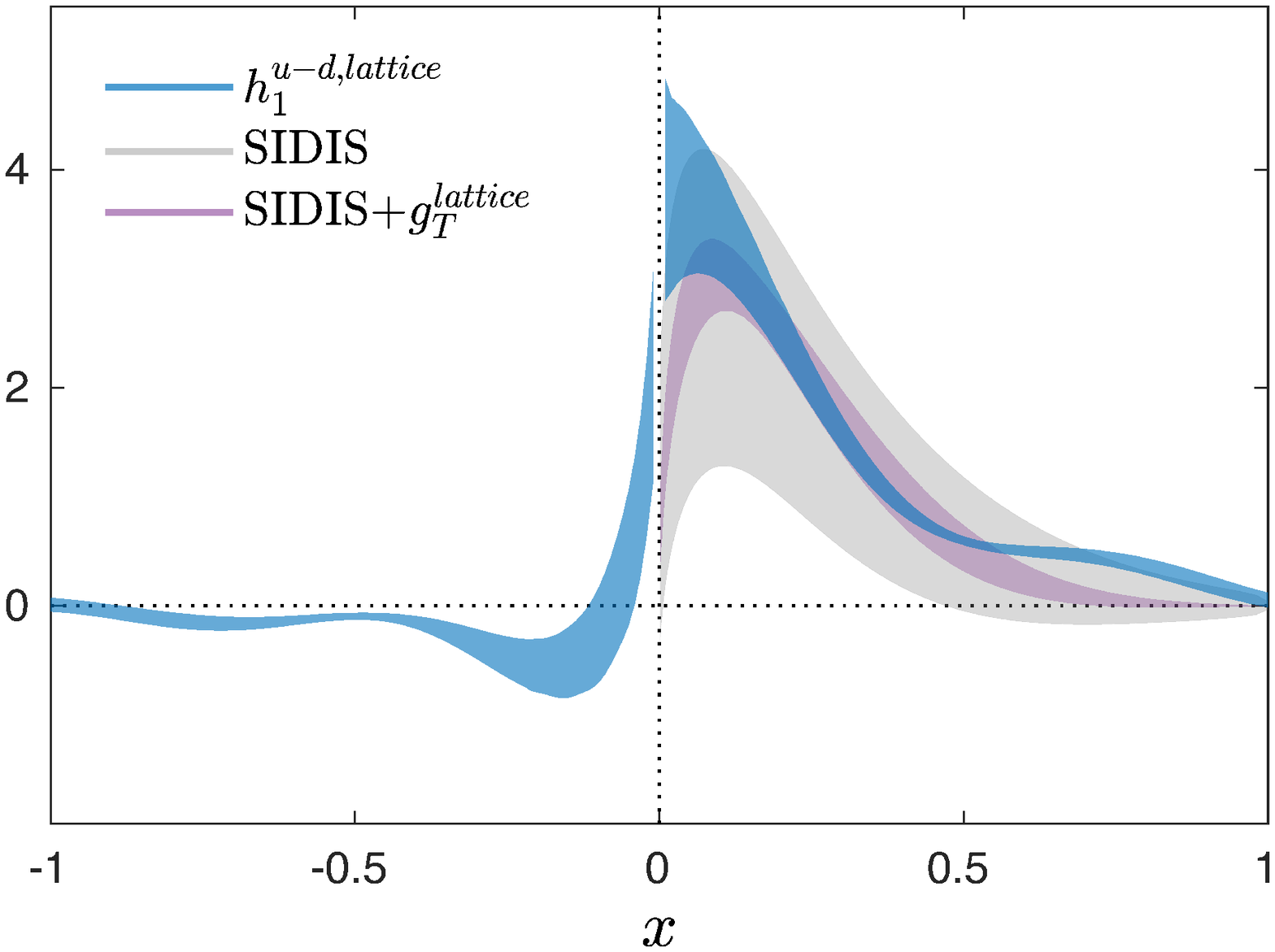}
   \end{minipage}
   \caption{Isovector unpolarized (left), helicity (middle) and transversity (right)  PDFs as a function of the momentum fraction $x$. The unpolarized and helicity PFDs are shown for  momenta $6\pi/L=0.83$~GeV (green band), $8\pi/L=1.11$~GeV (red band), and $10\pi/L=1.38$~GeV (blue band) and are compared with  results from  phenomenological analyses. The transversity is shown for $10\pi/L=1.38$~GeV (blue band) and compared to phenomenological fits  obtained using SIDIS data (gray) and SIDIS data constrained using the nucleon tensor charge computed in lattice QCD~\cite{Lin:2017stx}.}\vspace*{-0.7cm}
   \label{fig:PDFs}
\end{figure}

The starting point is the renormalized space-like matrix elements for boosted nucleon states 
\begin{equation}
  \tilde{F}_\Gamma(x,P_3,\mu) =2P_3
  \int_{-\infty}^{\infty}\frac{dz}{4\pi}e^{-ixP_3z}\,\langle P_3\vert\,\overline{\psi}(0)\, \Gamma W(0,z)\, \psi(z)|\,P_3\rangle |_{\mu}\quad,
\end{equation}
where $W$ is a Wilson line. The quasi-PDF $\tilde{F}_\Gamma$ is then matched using
\be
\tilde{F}_\Gamma(x,P_3,\mu)=\int_{-1}^1
\frac{dy}{|y|} \, C\left(\frac{x}{y},\frac{\mu}{yP_3}\right)\, {F}_\Gamma(y,\mu) +{\cal{O}}\left(\frac{m_N^2}{P_3^2},\frac{\Lambda^2_{\rm QCD}}{P_3^2}\right)
\ee
the light-cone PDF $F_\Gamma(x,\mu)$. We extract  the unpolarized, helicity and transversity PDFs when we take   for $\Gamma=\gamma_0, \gamma_i\gamma_5$ and $\sigma_{i,j}, i\ne j$, respectively. Results using an $N_f=2+1+1$ twisted mass fermion ensemble on the three isovector PDFs are shown in Fig.~\ref{fig:PDFs}. For details we refer to Refs.~\cite{Alexandrou:2018pbm,Alexandrou:2018eet}.
\begin{figure}[ht!]
   \begin{minipage}{0.4\linewidth}
     \includegraphics[width=\linewidth]{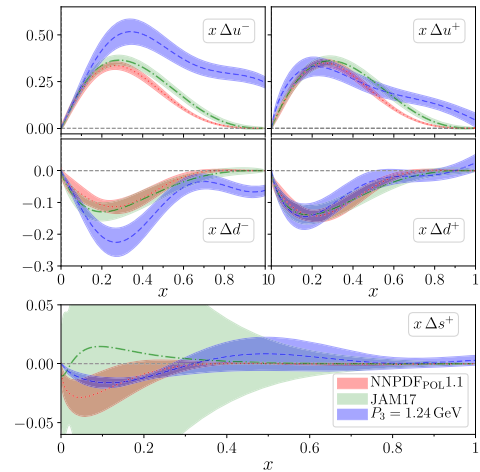}
     \caption{The up (top), down (middle) and strange (bottom) quark  helicity PDF (blue) versus $x$ in the $\overline{\rm MS}$ scheme at 2~GeV computed using an $N_f=2+1+1$ TMF ensemble with pion mass $m_\pi=260$~MeV~\cite{Alexandrou:2020uyt}. We compare with results from phenomenological analyses by JAM17 (green~\cite{Ethier:2017zbq} and NNPDF$_{\rm POL}1.1$ (red)~\cite{Nocera:2014gqa}.\label{fig:uds PDFs}}
   \end{minipage}\hfill
   \begin{minipage}{0.57\linewidth} To determine the flavor decomposition of PDFs one needs to compute, besides the isovector, the  isoscalar PDFs. The latter as well as  the strange PDFs involve disconnected contributions. These are computed by extending the formalism for local operators to extended operators. The quark disconnected loop is given by
     \be
     {\cal{L}}(t_{\rm ins},z) =\sum\limits_{\vec{x}_{\rm ins}} {\rm Tr}\left[D_q^{-1}(x_{\rm ins};x_{\rm ins}+z)\gamma^3\gamma^5 W(x_{\rm ins},x_{\rm ins}+z)\right],\nonumber
     \ee
     where we took the Wilson line along the z-direction. The helicity up, down and strange PDFs are shown in Fig.~\ref{fig:uds PDFs}~\cite{Alexandrou:2020uyt,Alexandrou:2021wzv}.
     
     The quasi-PDF approach can be extended to study generalized parton distributions (GDPs), where momentum transfer is involved. One computes the nucleon matrix element
     \beq &&h_\Gamma(z,\tilde{\xi},Q^2,P_3) =\nonumber\\
     &&\langle N(P_3\hat{e}_z+\vec{Q}/2)\vert\,\overline{\psi}(z)\, \Gamma W(z,0)\, \psi(0)|\,N(P_3\hat{e}_z-\vec{Q}/2)\rangle, \nonumber
     \eeq
     where the quasi-skewness
        $\tilde{\xi}=-\frac{Q_3}{2P_3}=\xi+{\cal{O}}(\frac{1}{P_3^2})$ with $\xi$ the skewness~\cite{Alexandrou:2020zbe}.
      \end{minipage}
\end{figure}

\section{Conclusions}
Moments of PFDs can be extracted precisely making these quantities part of  the precision era of lattice QCD. From these moments  we can extract a lot of interesting physics and also reconstruct the PDFs, as was demonstrated for the pion\cite{Alexandrou:2021mmi}.
 New theoretical approaches (quasi-distributions, pseudo-distributions, current-current correlates, etc) now allow the direct computation of PDFs within lattice QCD  yielding very promising results including  calculations using simulations with physical pion mass.
 The calculation  of sea quark contributions to PDFs is shown to also be feasible providing valuable input e.g. for the determination of the strange helicity PDF.
 Exploratory studies of isovector GPDs are also under way, see e.g. recent studies by ETMC~\cite{Alexandrou:2020zbe,Alexandrou:2021bbo,Alexandrou:2021lyf}.
The way forward includes studying the 
  continuum limit and the volume dependence and developing noise reduction techniques to reach larger boosts. Extending these approaches to  
 twist-3 PDFs\cite{Bhattacharya:2020cen}, TMDs, and other hadrons is also in the pipeline.

 \vspace*{0.5cm}
 
 \noindent
{\bf Acknowledgments}\\
   The work is financially supported by the Cyprus Research and Innovation Foundation under contract numbers POST-DOC/0718/0100 and COMPLEMENTARY/0916/0015, the project ``Nucleon parton distribution functions using Lattice Quantum Chromodynamics"  funded by the University of Cyprus and  the Horizon 2020 research and innovation program
of the European Commission under the Marie Sk\l{}odowska-Curie grant agreement No 642069.
We gratefully acknowledge the Gauss Centre for Supercomputing e.V. (www.gauss-centre.eu)
for funding the project pr74yo by providing computing time on the GCS Supercomputer SuperMUC
at Leibniz Supercomputing Centre (www.lrz.de).
Results were obtained using Piz Daint at Centro Svizzero di Calcolo Scientifico (CSCS),
via the project with id s702.
We thank the staff of CSCS for access to the computational resources and for their constant support.
This work also used computational resources from Extreme Science and Engineering Discovery Environment (XSEDE),
which is supported by National Science Foundation grant number TG-PHY170022.
We acknowledge Temple University for providing computational resources, supported in part
by the National Science Foundation (Grant Nr. 1625061) and by the US
Army Research Laboratory (contract Nr. W911NF-16-2-0189).
This work used computational resources from the John von Neumann-Institute for Computing on the Jureca system at the research center
in J\"ulich, under the project with id ECY00.


\end{document}